\documentclass[prl,twocolumn,aps,groupedaddress,showpacs]{revtex4}
\usepackage{amsmath,amsfonts,amssymb}
\usepackage{wrapfig}
\usepackage{graphicx}
\usepackage{bbm}
\usepackage{graphics}

\newcommand{\ez}{\epsilon_{z}}

\newcommand{\upp}{\hspace{-0.2pt}\uparrow}

\newcommand{\downn}{\hspace{-0.2pt}\downarrow}

\newcommand{\mean}[1]{\langle#1\rangle}

\newcommand{\ket}[1]{\left|#1\right\rangle }

\begin{document}
\title{Solid-state circuit for spin entanglement generation and purification}
\author{J. M. Taylor$^1$, W. D\"ur$^{2,3}$, P. Zoller$^{2,3}$, A. Yacoby$^{1,4}$, C. M. Marcus$^1$, and M. D. Lukin$^1$}
\affiliation{
$^1$ Department of Physics, Harvard University, Cambridge,
Massachusetts 02138, USA \\
$^2$  Institut f\"ur Theoretische Physik, Universit\"at Innsbruck, Technikerstra{\ss}e 25, A-6020 Innsbruck, Austria \\
$^3$  Institut f\"ur Quantenoptik und Quanteninformation der \"Osterreichischen Akademie der Wissenschaften, Innsbruck, Austria \\
$^4$ Department of Condensed Matter Physics, Weizmann Institute
of Science, Rehovot 76100, Israel}

\begin{abstract}
We show how realistic charge manipulation
and measurement techniques, combined with the exchange interaction,
allow for the robust generation and purification of four-particle
spin entangled states in electrically controlled semiconductor quantum dots.
The generated states are immunized
to the dominant sources of noise via a dynamical decoherence-free subspace;
all additional errors are corrected by a purification protocol. This approach may find application in quantum computation, communication, and metrology.
\end{abstract}
\maketitle

Spin entangled states are a basic resource for quantum information
processing, including quantum communication,
teleportation, measurement-based quantum computation~\cite{gottesman99} and quantum-based metrology. EPR pairs exemplify
spin entangled states, contributing both to theoretical insight into
the nature of entanglement, and
to experimental proofs of Bell's inequalities~\cite{aspect82}. In addition, entangled pairs are
a fundamental component in scaling up quantum computers, by connecting
small-scale processors in a quantum network. EPR pair generation and
purification is traditionally discussed in the context of long distance
quantum communication via photons in a quantum repeater setup. 
In the presence
of errors in noisy communication channels, robust generation of high-fidelity
EPR pairs can be achieved via purification \cite{bennett96,deutsch96,briegel98},
where a single high-quality pair is distilled in a probabilistic manner
from many low-fidelity singlets. In
a solid state environment, these ideas remain relevant, for example
for spin-based qubits in quantum dots, from the perspective of connecting
``distant'' parts of mesoscopic circuits, as well as from the
more fundamental perspective of protection of entanglement in a 
complex environment.
This letter develops a protocol for generation and purification
of electron spin-based EPR pairs in mesoscopic circuits, which builds
directly on emerging experimental techniques, and is tailored to the
specific decoherence mechanisms in a semiconductor environment.

\begin{figure}

\includegraphics[%
  width=3in]{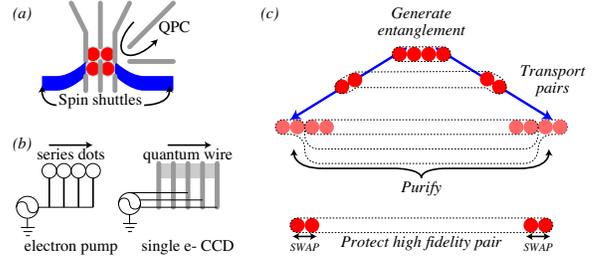}

\caption{
(a) Schematic outline of a node, as it might be implemented
in an gate-defined quantum dots (red). A nearby quantum
point contact (QPC) measures charge,
gates (gray) are pulsed for state generation and control, and spin transport channels (blue) allow the entangled
state to be sent to distant locations. (b) Schematic outline of two
spin transport channels: electron pump and single electron CCD. (c)
Overview of robust entanglement generation: generate entanglement
at a central node, transport the separated entangled states to end nodes, purify to remove noise encountered during transport and
due to memory-related errors, and protect pairs using dDFS techniques while
waiting for the next entangled state.
\label{f:overview}}
\end{figure}

We consider a setup consisting of an array of electrically gated quantum
dots (see Fig.~\ref{f:overview}), where electrons, with spin representing
the qubit, can be transported by applying appropriate gate voltages
(\cite{ono05,kielpinski:02}, Fig.~\ref{f:overview}b). In its simplest form,
a nonlocal EPR pair of electrons can be produced by local preparation
of a ground singlet state of two electrons in one of the quantum dots, splitting the pair
into two adjacent dots and shuttling the electrons to the end nodes.
A purification protocol corrects for qubit errors from the transport and storage. Our strategy is to develop such
a purification protocol on a more advanced level, where the qubits
are encoded in \emph{logical states} of a decoherence free subspace
(DFS) of two electrons, which from the beginning immunizes our logical
qubits against the dominant source of decoherence represented
by hyperfine interactions. Thus the goal is to produce local pairs
of \emph{logical entangled states}, represented by four entangled electrons,
transport logical pairs to the end nodes, and run an Oxford-type
\cite{deutsch96} \emph{purification protocol} on these
logical qubits that corrects all errors. 
We will show below that exchange interactions and
(partial) Bell measurements for the physical qubits are sufficient
to implement this protocol.  We remark that the required physical resources are already available at present in the lab.

\paragraph{Choice of encoded states} 

We begin by describing a specific encoding that allows suppression of the dominant error mechanism.  We focus on
hyperfine effects as theory and experiment
have demonstrated their detrimental effect on electron
spin coherences (dephasing), with $T_{2}^{*}\sim10$ ns \cite{bracker04,petta05}, while spin-orbit-phonon and other spin-flip processes (relaxation)
are observed to enter only for times on the order of 1 ms in the presence
of a large magnetic field \cite{hanson03b,golovach04}. Given the long
correlation time of the electron spin-nuclear spin interaction \cite{mehring76,merkulov:02},
storing entanglement in the logical states of a DFS with total $S_z$ quantum number $m_s=0$, $\ket{0_{L}}=(\ket{\upp\downn}-\ket{\downn\upp})/\sqrt{2}$ and $\ket{1_{L}}=(\ket{\upp\downn}+\ket{\downn\upp})/\sqrt{2}$,
allows for suppression of such dephasing
by repeatedly exchanging the two electrons~\cite{wu02}. 
The four-particle entangled state
\begin{equation}
\ket{\phi^{+}}=\ket{0_{L}}\ket{0_{L}}+\ket{1_{L}}\ket{1_{L}}=\ket{\upp\downn\upp\downn}+\ket{\downn \upp\downn\upp}
\end{equation}
takes full advantage of these properties,  suppressing phase noise. This combination of subspace choice and exchanging electrons
corresponds to a dynamical DFS (dDFS) and is a de facto implementation of Carr-Purcell spin echo in the DFS.
We show below that using dDFS, memory and transport
errors will be dominated by spin-flip terms, an improvement
of order $10^{5}$ over hyperfine-related noise.  Errors in the dDFS procedure, and spin-flip errors, are so far uncorrected.
Starting with several copies of the entangled state $\ket{\phi^{+}}$, our purification
protocol corrects for spin-flip errors entirely, by detection of the
total $m_{s}$ quantum number of the states, while it corrects for
phase errors by analogy to the protocol of Ref.~\cite{deutsch96}.

We now consider the ingredients and recipe for EPR generation and purification in the DFS: (I) charge manipulation and measurement
techniques for performing exchange gates ($U_{AB}(\phi)$), singlet
generation, and partial Bell state measurements $M_{AB}$; (II) the
dynamical DFS's properties with regards to different noise sources,
its behavior during storage (memory) and transport, to show suppression
of better than $10^{5}$ for low frequency phase noise; (III) a purification
protocol that works in the encoded space and corrects for arbitrary
errors, using only the partial Bell state measurement and exchange
gate described in (I).

\paragraph{I. Charge manipulation and measurement}

We suggest an implementation of the necessary resources for each node: exchange gate, singlet generation, and partial 2 electron Bell state measurement.
In principle, other techniques could be
used to generate the same set of operations.

The Loss-Divincenzo exchange gate \cite{loss98} between two electrons
in separate dots, $A\ {\rm and}\  B$, is defined as $U_{AB}(\phi)=\exp(-i\phi\vec{S}^{A}\cdot\vec{S}^{B})$;
for example, $U(\pi/8)$ is $\sqrt{{\rm SWAP}}$. By control of the tunnel coupling ${\textrm{T}_c}$ between
$A\ {\rm and}\  B$, or by changing their relative bias, arbitrary $\phi$ may be achieved. It requires only pulsed-gate manipulation, i.e., it relies on
charge control. 

\begin{figure}

\includegraphics[%
  width=3in]{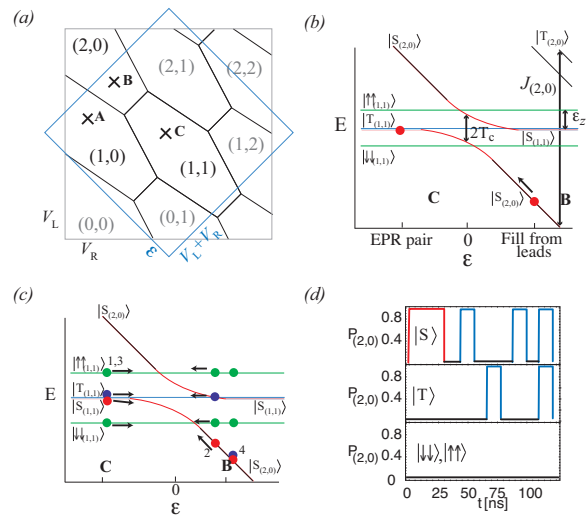}

\caption{(a) Stability diagram of a double dot system, with gate voltages
for left and right dots, $V_{L},V_{R}$ and alternative axes of detuning
and total voltage, $\epsilon,V_{L}+V_{R}$. The three positions A, B, and C are marked
(b) Energy level diagram for detuning
between B and C. 
An external magnetic field
Zeeman ($\ez$) splits the triplet levels of the (1,1) configuration;
tunnel coupling leads to an avoided crossing at $\epsilon=0$; by
starting in $\ket{S_{2,0}}$ at B and adiabatically changing $\epsilon$
to C a separated singlet is generated. (c) The four stage measurement
procedure, as described in the text. (d) Example signals of the measurement for four possible initial states, as labeled; section in red occurs only for $\ket{S}$; blue only for $\ket{S},\ket{T_0}$. \label{f:measurement}}

\end{figure}

Singlet states of double dots may be created using the large exchange
splitting of single dots. For a double-dot system,
starting in the (1,0) stability island (Fig.~\ref{f:measurement}a) resets the state of the double-dot
(position A); changing configuration to the (2,0) stability
island (position B) and coupling to the leads results in
a singlet state of (2,0) $(\ket{S_{(2,0)}})$ if the single dot exchange
is large, $J_{(2,0)}\gg k_{b}T$, which prevents filling of the triplet
states.
We remark that this is the only strict temperature requirement in this paper.
Adiabatically changing the bias of the double dot system to the (1,1)
stability island (position C, Fig.~\ref{f:measurement}b) results in adiabatic passage
of the (2,0) singlet to the (1,1) singlet $(\ket{S_{(1,1)}}=(\ket{\upp\downn}-\ket{\downn\upp})_{(1,1)}/\sqrt{2})$.
If this is accomplished much faster than dephasing mechanisms, the
(1,1) singlet can be prepared with high fidelity. 
Assuming a linear ramp of detuning with a time $\tau$, the probability of error
goes as $\pi (\tau/T_2^*)^2 (10^{-2} + (\hbar/\epsilon_z T_2^*)^{2})$; for $\tau=1$ ns and $T_2^* = 10$ ns, the fidelity is $>0.99$.

In addition, we can exploit the double-dot system to make a partial
Bell measurement that leaves a logical subspace of our system untouched (up
to a correctable phase). To achieve this for two spins in separated but adjacent
quantum dots, the detuning is adiabatically changed from position C
to position B. Only the singlet ($\ket{S}$) transfers; waiting
a time $t_{1}$ in this configuration allows a charge measurement
to distinguish between (2,0) ($\ket{S}$ result) and (1,1) (one of
three triplet states). Adiabatically returning to C and waiting
a time $T_{2}^{*}$ switches the singlet and $m_{s}=0$ triplet ($\ket{T_{0}}$)
states with probability 1/2. Going again to B, if the triplet switched to
$\ket{S}$, it transfers to (2,0), producing a noticeable
charge signal. Repeating this process $k$ times can generate, with probability
$1-1/2 \sqrt{k}$, a charge signal for the $m_{s}=0$ subspace; the total
time for charge measurement is $t_{M}/2\simeq k(T_{2}^{*}+t_m)+t_{1}$, where $t_m$ is the time to make a single charge measurement. In our present implementation, long $t_m$ may be the main limitation for the purification protocol discussed below~\cite{foot1}.
The three results of measurement are (a) singlet, (b) $m_{s}=0$ triplet,
or (c) $|m_{s}|=1$. During this time, the $|m_{s}|=1$ states remain untouched
except for a phase; we now show how this measurement procedure, denoted
$M_{AB}$, can generate our desired entangled state, $\ket{\phi^{+}}$,
and adjust the phase for such a state.

Starting with four dots (1--4) (Fig.~\ref{f:overview}c), we prepare singlets in 12 and 34;
this initial state is $\ket{S}_{12}\ket{S}_{34}$. Applying $M_{23}$
and keeping only the $|m_{s}|=1$ result (occurring with probability
1/2) yields the state $\ket{\phi^{+}}$. To correct the accumulated phase error on it, 
we use a sequence: {[}wait($t_{M}/4$), SWAP$_{12}$, SWAP$_{34}$,
$M_{23}$, SWAP$_{12}$, SWAP$_{34}$, wait($t_{M}/4$){]}, which we now study.

\paragraph{II. Dynamical DFS}

We examine the dynamical DFS in detail, with a general noise formulation.
While we focus on hyperfine terms, other low
frequency noise will be similarly corrected. To be specific, we assume
a phase noise term $\eta(t)$ acts on electron spins, characterized by a power spectrum,
$S(\omega)$ of integrated power $(T_2^*)^2$ with a (possibly polynomial) high frequency cutoff at
$\gamma\ll1/T_{2}^{*}$. For example, the
hyperfine interaction in quantum dots, with long-time scale non-Markovian dynamics, is well described by this process~\cite{merkulov:02}. 

In a frame rotating with external magnetic field (which also defines
up and down spin), the phase term acts on a spin state as $\ket{\upp}\pm\ket{\downn}\rightarrow\ket{\upp}\pm e^{-i\int_{0}^{t}\eta(t')dt'}\ket{\downn}$.
Using two electron spins in separate, adjacent dots to create the
encoded space, $\ket{0_{L}},\ket{1_{L}}$, this action may be represented
by a stochastic evolution operator, $U(t,0)=e^{-i\int_{0}^{t}\eta(t')dt'\sigma_{x}^{L}}$,
where $\sigma_{x}^{L}$ is a Pauli matrix for the encoded space, i.e.,
flips the logical bit. As the dots are adjacent, they may easily be
SWAPed. 
The pulse sequence {[}wait($\tau/4$), SWAP, wait($\tau/2$),
SWAP, wait($\tau/4$){]} gives 
a reduced power spectrum,
\begin{equation}
S_{dDFS}(\omega)=S(\omega)\frac{256}{\tau^{2}\omega^{2}}\cos^{2}(\frac{\tau\omega}{8})\sin^{6}(\frac{\tau\omega}{8})~.\end{equation}
 For frequencies below $1/\tau$, $S_{dDFS}(\omega)\simeq S(\omega)\frac{\tau^{4}\omega^{4}}{1024}$;
if the dominant noise mechanism has only low frequency components
(such as hyperfine terms) the suppression can be dramatic. For SWAP operations performed by use of exchange gates, the gate
must be performed in a time $\tau_{ex}\ll
T_{2}^{*}$;
with physical exchange of electrons, e.g., by use of an auxiliary
dot, this requirement is relaxed.

The DFS also reduces phase errors incurred during transport of the
electron spins. For example, two electrons forming a logical
state are moved sequentially through the same channel (i.e., same
series of quantum dots) with a separation time $\tau_{T}\ (\approx4\sigma/v$,
where $\sigma$ is the lateral radius of each dot, and $v$ the average
velocity of transport). Replacing $\eta(t)$ with $\eta(x,t)$, we set $\mean{\eta(x,t)\eta(x',t')}=C(|x-x'|)\int_{-\infty}^{\infty}S(\omega)e^{i\omega(t-t')}d\omega$
for transport through a series of quantum dots, where $C(x)=e^{-x^{2}/2\sigma^{2}}$.
The resulting spectral function is
\begin{equation}
S_{T}(\omega)=\int_{-\infty}^{\infty}S(\omega-\nu)\sin^{2}[(\omega-\nu)\frac{\tau_{T}}{2}]\frac{e^{-(\frac{\tau_{T}}{4})^{2}\nu^{2}/2}}{\sqrt{2\pi(4/\tau_{T})^{2}}}\  d\nu~. \nonumber
\end{equation}
which has a suppression of noise with frequencies $\ll1/\tau_{T}$.
In particular, $S_{T}(\omega)$ corresponds to $S(\omega)$ integrated
in a window of size $\sim1/\tau_{T}$ and suppressed at low frequencies
by $\tau_{T}^{2}\omega^{2}/8$. 

Considering practical parameters, we set $\gamma=\gamma_{dd}=1$ ms$^{-1}$,
$T_{2}^{*}=10$ ns, and use $S(\omega)=e^{-\frac{\omega^{2}}{2\gamma^{2}}}/(T_2^* \sqrt{2\pi\gamma^{2}})$.
For states stored in the dDFS with a cycle time $\tau$, after one
cycle the probability of error is $p_{{\rm err}}=\frac{3}{2^{12}}\frac{\gamma^{4}\tau^{6}}{(T_{2}^{*})^{2}}$.
Transporting through $n=L/v$ quantum dots,
we find the probability of a phase error occurring for the encoded
states is $p_{{\rm err},T}(n)\approx\sqrt{\frac{\pi}{128}}\left(\frac{\gamma}{T_{2}^{*}}\right)^{2}\tau_{T}^{4}n
$. 
%If using resonant exchange interaction, $\tau_{ex} \ll T_{2}^{*}$.  
Even for cycle and transport times ($\tau,\tau_{T})$ approaching
$T_{2}^{*}$, phase errors due to low frequency terms occur with rates
much slower than milliseconds, indicating a suppression of more than $10^5$. Thus the dynamical DFS technique provides
a powerful quantum memory and low-error transport channel, limited
by errors in SWAP operations and spin-flip processes.

\paragraph{III. Purification}

We now introduce a purification protocol for encoded entangled states
that can remove all remaining errors, based on partial Bell measurement
and exchange gates. Errors during the generation, transport, and storage
processes can lead to (i) errors within the $\{|0_{L}\rangle,|1_{L}\rangle\}$
logical subspace, and (ii) population of states $|2_{L}\rangle=1/\sqrt{2}(\ket{\uparrow\upp}+\ket{\downn\downn})$,
$|3_{L}\rangle=1/\sqrt{2}(\ket{\upp\upp}-\ket{\downn\downn})$
outside the logical subspace. Both kind of errors reduce the fidelity
of the encoded entangled state $|\phi^{+}\rangle_{A_{1}A_{2}B_{1}B_{2}}$
and need to be corrected. We introduce a purification protocol that
completely corrects arbitrary strength errors of type (ii), and corrects
for errors of type (i) that occur with probability less than 1/2.

We start by reviewing the measurement scheme (Fig.~\ref{f:measurement}c,d),
which has three possible outcomes: (a) $P_{S}$: measure $|S\rangle$,
state after measurement is $|S\rangle$; (b) $P_{T_{0}}$: measure
$|T_{0}\rangle=1/\sqrt{2}(\ket{\upp\downn}-\ket{\downn\upp})$, state
is $|S\rangle$; (c) $P_{|m_{s}|=1}$: state is coherently projected
into the two--dimensional subspace spanned by $\{\ket{\upp\upp},\ket{\downn\downn}\}$.
Consider the following sequence of measurements of this type with
indicated results: $O_{A}^{(0)}=P_{S}^{(A_{1}'A_{2}')}P_{|m_{s}|=1}^{(A_{1}A_{1}')}P_{|m_{s}|=1}^{(A_{2}A_{2}')},O_{A}^{(1)}=P_{T_{0}}^{(A_{1}'A_{2}')}P_{|m_{s}|=1}^{(A_{1}A_{1}')}P_{|m_{s}|=1}^{(A_{2}A_{2}')}$.
The action of $O_A^{(k)}$ on logical basis states is given by 
\begin{equation}
O_{A}^{(k)}|i_{L}\rangle_{A_{1}A_{2}}|j_{L}\rangle_{A_{1}'A_{2}'}=P_{\{0,1\}}|i_{L}\oplus j_{L}\oplus k\rangle_{A_{1}A_{2}}|0_{L}\rangle_{A_{1}'A_{2}'}, \nonumber
\end{equation}
 where the projector $P_{\{0,1\}}$ indicates that, in both $A,A'$, all components
outside the $\{|0_{L}\rangle,|1_{L}\rangle\}$ subspace 
are projected
out. The measurement sequence can thus be used to detect all errors
of type (ii), while the operations $O_{A}^{(k)}$ act
within the 
logical subspace similarly
to a CNOT operation ($\oplus$ denotes bit-wise addition modulo 2).

Consider a mixed state $\rho_{A_{1}A_{2}B_{1}B_{2}}$ resulting from
imperfectly distribution of $|\phi^{+}\rangle$.
We decompose $\rho$ into three terms, $\rho=\rho(\vec{x})+\rho_{\textrm{od}}+\rho_{R}$.
We have $\rho(\vec{x})=x_{0}|\phi^{+}\rangle\langle\phi^{+}|+x_{1}|\phi^{-}\rangle\langle\phi^{-}|+x_{2}|\psi^{+}\rangle\langle\psi^{+}|+x_{3}|\psi^{-}\rangle\langle\psi^{-}|$,
where $\ket{\phi^{\pm}}=(\ket{0_{L}0_{L}}\pm\ket{1_{L}1_{L}})/\sqrt{2}$
and $\ket{\psi^{\pm}}=(\ket{0_{L}1_{L}}\pm\ket{1_{L}0_{L}})/\sqrt{2}$
are the logical Bell states. All off-diagonal elements in the Bell basis ($\rho_{\textrm{od}}$) and terms containing $\{\ket{2_L},\ket{3_L}\}$ ($\rho_R$) are made irrelevant by the protocol.

Given two mixed states $\rho_{A_{1}A_{2}B_{1}B_{2}}\otimes\rho'_{A'_{1}A'_{2}B'_{1}B'_{2}}$,
described by $\vec{x}$ and $\vec{x}'$ (and the irrelevant $\rho_{\textrm{od}}+\rho_R$) respectively, the following sequence of
\emph{local} operations obtains with certain probability a state
with higher fidelity and hence purifies the state: (i) partial depolarization
of $\rho$ using, with probability $p=1/2$, SWAP$_{A_{1}A_{2}}\otimes$
SWAP$_{B_{1}B_{2}}$ or identity, and similarly for
$\rho'$; (ii) exchange gates $U(\pi/8)_{A_{1}A_{2}}\otimes U(-\pi/8)_{B_{1}B_{2}}$
at $\rho_{A_{1}A_{2}B_{1}B_{2}}$ (and same for $\rho')$; (iii) Sequence
of measurements $O_{A}^{(k)}$, $O_{B}^{(l)}$; keep state $\rho_{A_{1}A_{2}B_{1}B_{2}}$
only if $k=l$, i.e. the results in final measurement coincide in
$A$ and $B$.

The effect of (i) is to erase off--diagonal terms of the form $|\phi^{\pm}\rangle\langle\psi^{\pm}|$
which may contribute to the protocol. The operation in (ii) exchanges
logical states $|\psi^{+}\rangle\leftrightarrow|\psi^{-}\rangle$
while keeping $|\phi^{\pm}\rangle$ invariant. Finally (iii) realizes
--in addition to the projection into the $\{|0_{L}\rangle,|1_{L}\rangle\}$
logical subspace in $A$ and $B$ which erases all terms $\rho_{R},\rho'_{R}$--
a purification map. In particular, we find that the remaining off
diagonal elements do not contribute and the action of the protocol
can be described by the non-linear mapping of corresponding vectors
$\vec{x}$, $\vec{x}'$. The resulting state is of the form $\rho(\vec{y})+\tilde{\rho}_{\textrm{od}}$
(note that $\rho_{R}=0$), where \begin{eqnarray}
y_{0}  =  (x_{0}x'_{0}+x_{2}x'_{2})/N &,&
y_{1}  =  (x_{1}x'_{1}+x_{3}x'_{3})/N~,\nonumber \\
y_{2}  =  (x_{1}x'_{3}+x_{3}x'_{1})/N &,&
y_{3}  =  (x_{0}x'_{2}+x_{2}x'_{0})/N~,\end{eqnarray}
 and $N=x_{0}x'_{0}+x_{2}x'_{2}+x_{1}x'_{1}+x_{3}x'_{3}+x_{1}x'_{3}+x_{3}x'_{1}+x_{0}x'_{2}+x_{2}x'_{0}$
is the probability of success of the protocol. This map is equivalent
(up to a reduced success probability by a factor of 1/8) to the purification
map obtained in Ref. \cite{deutsch96} for non--encoded Bell states.
It follows that iteration of the map --which corresponds to iteratively
applying the purification procedure (i-iii) to two identical copies
of states resulting from successful previous purification rounds--
leads to a (encoded) maximally entangled state $|\phi^{+}\rangle$.
That is, the map has $\vec{y}=(1,0,0,0)$ as attracting fixed point
whenever $x_{0}>x_{1}+x_{2}+x_{3}$ . We emphasize that \emph{all}
errors leading outside the logical subspace (in particular all spin
flip errors), independent of their probability of occurrence, can
be corrected. This implies that even states with a very small fidelity
$F$ can be purified, provided that errors within the logical subspace
do not exceed probability 1/2. 

Additionally, since the resulting maps
are identical to those of \cite{deutsch96}, the purification protocol
shows a similar robustness against noise in local control operations.
That is, errors of the order of several percent in local control operations
can be tolerated while still leading to purification. We also remark
that methods such as (nested) entanglement pumping can be applied
\cite{dur03}, which significantly reduces the required number nodes, and may be used for a full quantum repeater protocol.

While we have focused on gate controlled quantum dots, these ideas may find implementation in electro-optically manipulated small arrays of self-assembled quantum dots~\cite{krenner05}. In general, the prescription for entanglement generation in solid-state environments
we describe here could also be followed in other solid-state systems
such as superconductor-based qubit designs~\cite{wallraff04}. We anticipate that such
long-range entangled state generation will have wide application,
in scalable quantum computer architectures, 
in tomography based on entangled states, and in the fundamental physics
of noise in solid-state environments.

We gratefully acknowledge helpful conversations with J. Petta and A. Johnson. 
The work at Harvard was supported by ARO, NSF,
Alfred P. Sloan Foundation, and David and Lucile Packard
Foundation. The work at Innsbruck was supported by the \"OAW through project APART (W.D.), the European Union, Austrian Science Foundation, and the DFG.

\end{document}